\documentclass{article}
\def\giorno{21/01/2020}

\usepackage{color}
\usepackage{amsmath,amsfonts,amssymb}

\def\a{\alpha}
\def\b{\beta}
\def\ga{\gamma}
\def\de{\delta}   

\def\vphi{\varphi}
\def\la{\lambda}

\def\s{\sigma}

\def\om{\omega}

\def\vphi{\varphi}

\def\C{{\bf C}}

\def\G{{\cal G}}

\def\Ga{\Gamma}
\def\De{\Delta}
\def\La{\Lambda}

\def\L{\mathcal{L}}

\def\R{{\bf R}}

\def\T{{\rm T}}

\def\pa{\partial}

\def\d{{\rm d}}       
\def\w{\wedge}

\def\o+{\oplus}
\def\xd{{\dot x}}

\def\ss{\subset}
\def\sse{\subseteq}

\def\<{\langle}
\def\>{\rangle}

\def\({\left(}
\def\){\right)}
\def\[{\left[}
\def\]{\right]}
\def\=#1{\bar #1}
\def\wt#1{\widetilde #1}
\def\.#1{\dot #1}
\def\^#1{\widehat #1}

\def\"#1{\ddot #1}

\def\interno{\hskip 2pt \vbox{\hbox{\vbox to .18
truecm{\vfill\hbox to .25 truecm
{\hfill\hfill}\vfill}\vrule}\hrule}\hskip 2 pt}

\def\eeq{\end{equation}}
\def\beq{\begin{equation}}

\def\beql#1{\begin{equation} \label{#1}}

\def\eqref#1{(\ref{#1})}

\def\EOR{ \hfill $\odot$ \medskip}

\def\symmref{AVL,CGbook,KrV,Olv1,Olv2,Ste}
\def\MuRaltri{MuRom3,MuRom4,MuRom4b,MuRom5,MuRom6,MuRom7,MuRom11,MuRom11b,MuRom12,MuRom14}

\def\mapright#1{\smash{\mathop{\longrightarrow}\limits^{#1}}}
\def\mapdown#1{\Big\downarrow\rlap{$\vcenter{\hbox{$\scriptstyle#1$}}$}}

\def\interno{\hskip 2pt \vbox{\hbox{\vbox to .18
truecm{\vfill\hbox to .25 truecm
{\hfill\hfill}\vfill}\vrule}\hrule}\hskip 2 pt}

\begin{document}

\title{On the Geometry of Twisted Symmetries: \\ Gauging and Coverings}

\author{D. Catalano Ferraioli\thanks{Instituto de Matem\'atica e Estat\'istica - Universidade
Federal da Bahia, Campus de Ondina, Av. Adhemar de Barros, S/N, Ondina
- CEP 40.170.110 - Salvador, BA - Brazil; {\it e-mail:} {\tt diego.catalano@ufba.br}; {\it ORCID:} {\tt 0000-0002-5684-0493}} \ and G. Gaeta\thanks{Dipartimento di Matematica, Universit\`a degli Studi di
Milano, via Saldini 50, 20133 Milano (Italy) \& SMRI, 00058 Santa Marinella (Italy); {\it e-mail:} {\tt giuseppe.gaeta@unimi.it}; {\it ORCID:} {\tt 0000-0003-3310-3455}}}

\date{\giorno}

\maketitle

\begin{abstract} \noindent

We consider the theory of \emph{twisted symmetries} of
differential equations, in particular $\la$ and $\mu$-symmetries,
and discuss their geometrical content. We focus on their
interpretation in terms of gauge transformations on the one hand,
and of coverings on the other one.

\end{abstract}

{} \hfill {\it Dedicated to Josiph Krasil'shich on his 70$^{th}$ anniversary}

\section{Introduction}

The Geometry of Differential Equations has been a constant topic
in the research by Josiph Krasil'shich, and a great deal of this
has been devoted to (the symmetry approach to) the study of
\emph{symmetries of differential equations}.

These were first considered systematically by Sophus Lie, who laid
down the theory of point and contact symmetries. This theory was
later on generalized in several ways by many authors (including
JK). The basic idea by Lie is that once we know how the basic
(independent and dependent, possibly allowing first derivatives to
transform in a special way) variables transform, we also know how
higher derivative transform:this corresponds to the concept of
\emph{prolongation} of a vector field, which is thus lifted from
the phase  manifold $M$ to the associated \emph{jet bundle} $J^k
M$ or $J^\infty M$, of finite or infinite order \cite{\symmref}.

In most of the generalizations of Lie-point and contact
symmetries, this feature is preserved: one considers more general
types of vector fields in $M$ (e.g. generalized vector fields),
but the action these induce in $J^k M$ or $J^\infty M$ is
still obtained from the
action in $M$ by means of the standard \emph{prolongation}
operation -- and hence the standard prolongation formula.

There is, however, a class of generalizations for which this does
not hold true; these were first considered by Muriel and Romero
\cite{MuRom1,MuRom2} in the specific case of scalar
ODEs\footnote{From the point of view of the general theory built
afterwards, this is a degenerate case in many ways; which made not
so immediate to understand the underlying Geometry.}, and in this
case one speaks of \emph{$\la$-symmetries} or of
$C^\infty$-symmetries; in the general case they are known as
\emph{twisted symmetries} \cite{Gtwist1,Gtwist2}. For these, the
very prolongation operation is modified, so that the (twisted)
prolongation of a vector field in $M$ to $J^k M$ or $J^\infty M$
does not describe its action on (standard) derivatives. This
notwithstanding, twisted symmetries turn out to be ``as useful as
standard ones'' in reducing or solving nonlinear differential
equations (both ODEs and PDEs) and are thus of great interest both
from the abstract and geometrical point of view and from the
concrete and applicative one.

Over the years, we have (separately) worked on this topic, and
shown relations of it with two subjects which are also central in
the scientific interests of Josiph Krasil'shich; that is, the theory of
\emph{coverings} \cite{Kracov,KrV} on the one hand, and that of
\emph{gauge transformations} \cite{EGH,Nak,NS} one the other one (for the
relations with twisted symmetries, see \cite{CF,CFM1} and
\cite{Ggauge1,Ggauge2,Ggauge3} respectively).

The purpose of this paper is to review, and partially reconcile,
these two points of view on twisted symmetries, and their
relations with  relevant geometric structures.

\section{Symmetries of differential equations}

We assume the reader is familiar with symmetry of differential
equations; the purpose of this section is thus mainly to fix
notation.

We will consider differential equations\footnote{For the moment,
ODEs or PDEs will not make a difference, and differential
equations, are always possibly vector ones, i.e. systems;
similarly, functions are always possibly vector ones -- albeit in
some cases we will use vector indices explicitly to avoid possible
confusion.} with independent variables $x^i$ ($i=1,...,p$) and
dependent variables $u^a$ ($a = 1,..., q$); partial derivatives
will be denoted by $u^a_J$, where $J$ is a multi-index $J = \{ j_1
, ... , j_p \}$ of order $|J| = j_1 + ... + j_p$ and \beq u^a_J \
= \ \frac{\pa^{|J|} u^a}{\pa x_1^{j_1} ... \pa x_p^{j_p} } \eeq
(here and somewhere in the following we moved downstairs the
vector index of the $x$ for typographical convenience). We denote
by $u_{(k)}$ the set of all partial derivatives of order $k$, and
by $u_{[n]}$ the set of all partial derivatives of order $k \le
n$. We also denote by $\wt{J} = (J,i)$ the multi-index with
entries $\wt{j}_k = j_k + \delta_{ik}$.

The $x$ are local coordinates in a manifold $B$, while $u$ are
local coordinates in a manifold $U$; we consider the phase
manifold $M = B \times U$, which has a natural structure of bundle
$(M,\pi,B)$ over $B$ with fiber $U$.

We also associate to $M$ its Jet bundles $J^n M$, which associate
to any point $(x,u)$ the set of equivalence classes of
sections being mutually tangent of order $n$; these are
described in local coordinates by $(x,u,u_{(1)} , ... , u_{(n)}
)$. Note that $J^n M$ should be thought as equipped with a
\emph{contact structure}, generated by the contact forms \beq
\vartheta^a_J \ := \ \d u^a_J \ - \ u^a_{J,i} \, d x^i \ . \eeq

A (uni-valued) function $u = f(x)$ corresponds to a section
$\ga_f$ of $(M,\pi,B)$; this is just the graph of $f$, $$ \ga_f \
= \ \{ (x,u) \in B \times U \ : \ u = f(x) \} \ . $$ We will
denote the set of sections of $M$ as $\Sigma (M)$, and $\ga_f \in
\Sigma (M)$.

If we assign $u = f(x)$, we are implicitly assigning also all of
its derivatives; thus $\ga_f \in \Sigma (M)$ also identifies
prolongations (of any order) $\ga_f^{(n)} \in \Sigma (J^n M)$; in
multi-index notation, $$ \ga_f^{(n)} \ = \ \{ (x, u_{[n]}) \in J^n
M \ : \ u_J = (\pa_J f) (x) \ , \ |J| \le n \} \ . $$ These can be
thought of as sections of $(J^n M, \pi_n , B)$.


If we consider a differential equation\footnote{Note that by this we always
mean possibly a system of equations, ODEs or PDEs.} of order $n$, say \beql{eq:Delta} \Delta \ : = \ F^\ell
(x,u,u_{(1)} , ... , u_{(n)} ) = 0 \ \ \ (\ell = 1,...,L ) \eeq (we
always assume $F$ to be smooth in all of its arguments) this
identifies a manifold in $J^n M$, called the \emph{solution
manifold} $S_\De$; if $\De$ is non-degenerate, this is a manifold
of codimension $s$.

A function $u = f(x)$ is a solution to $\De$ if and only if
$$ \ga_f^{(n)} \ \ss \ S_\De \ \ss \ J^n M \ . $$

This also means that vector fields $Y$ in $J^n M$ which are both
tangent to $S_\De$ \emph{and} preserve the contact structure map
solutions into solutions.

The condition to preserve the contact structure can be stated more
precisely as follows: if $\Theta$ is the Cartan ideal generated by
the $\vartheta^a_J$, then $Y$ preserves the contact structure if
$$ \L_Y (\Theta ) \ \sse \ \Theta \ , $$ i.e. if for any $\om \in
\Theta$ we have $\L_Y (\om ) \in \Theta$. In view of the
properties of Cartan ideals, this is the case if and only if $\L_Y
(\vartheta^a_J) \in \Theta$, i.e. if and only if there are
functions $T^{aK}_{bJ} \in \C^\infty (J^n M , R)$ such that
$$ \L_Y (\vartheta^a_J ) \ = \ T^{aK}_{bJ} \ \vartheta^b_K \ . $$
By a standard computation, this is the case if and only if the
coefficients of the vector field
$$ Y \ = \ \xi^i \, \frac{\pa }{\pa x^i} \ + \ \psi^a_J \,
\frac{\pa }{\pa u^a_J} $$ satisfy the \emph{prolongation formula}
\beq \psi^a_{J,i} \ = \ D_i \psi^a_J \ - \ u^a_{J,k} \, (D_i \xi^k
) \ . \eeq

Note that -- setting $\psi^a_0 = \varphi^a$ -- this means that $Y$
is the prolongation of the vector field on $M$
$$ X \ = \ \xi^i \, \pa_i \ + \ \varphi^a \, \pa_a \ ; $$ this is
a well defined vector field in $M$ provided
$$ \xi^i \ = \ \xi^i (x,u) \ , \ \ \varphi^a  \ = \ \varphi^a
(x,u) \ ; $$ we will assume this to be the case\footnote{In other words,
here we are not considering contact or generalized vector fields and
symmetries.}, and in this case we also write
$$ Y \ = \ X^{(n)} $$ to emphasize that the vector field we are
considering in $J^n M$ is the \emph{prolongation} of the vector
field $X$ in $M$.

If such a vector field is tangent to $S_\De$, i.e.
\beq X^{(n)} \ : \ S_\De \to \T S_\De \ , \eeq we say that $X$ is a
Lie-point symmetry for $\De$. (More precisely, $X$ is then the
generator of a one-parameter local group of symmetries; but this
slight abuse of notation is commonplace in the literature, and we
will adhere to it.)

If $\De$ is written as in eq.\eqref{eq:Delta}, then the condition
that $X$ is a Lie-point symmetry can be expressed as
\beql{eq:Xsymm} X^{(n)} \[ F^\mu \]_{F=0} \ = \ 0 \ . \eeq

\medskip\noindent
{\bf Remark 1.}
Note that in \eqref{eq:Xsymm} we are only requiring the invariance of the level set ${\bf F} = 0$, \emph{not} of all the level sets ${\bf F} = {\bf c}$; in
the latter case, we would speak of \emph{strong symmetries}. \EOR


\section{Coverings and nonlocal symmetries}

We consider the notion of (first order) \emph{covering} of a
differential equation; here we discuss it in terms of coordinates,
for the sake of brevity; see \cite{Kracov,KrV} for an intrinsic
discussion.

Together with independent variables $x \in B$   and dependent ones
$u \in U$, with local coordinates respectively $(x^1,...,x^p)$ in
$B$ and $(u^1,...,u^q)$ in $U$, we consider auxiliary variables $w
\in W$, with $W$ a smooth manifold with local coordinates
$(w^1,...,w^r)$.

Then the system of $m$ equations \beql{eq:dede} \De \ := \ F^a (x,u,u_{(1)} , ... , u_{(n)} ) \ = \ 0 \ \ \ \ (a = 1,...,m) \eeq is augmented to a system $\wt{\De}$ of $m+s$ equations with a new set of $s = r \cdot p$ auxiliary first order equations
\beql{eq:w} w^\mu_i \ = \ H^\mu_i (x,u,w) \ . \eeq
This also means that the total derivative operators, which in $J^n M$ are
$$ D_i \ := \ \frac{\pa}{\pa x^i} \ + \ u^a_i \, \frac{\pa}{\pa u^a} \ + \ u^a_{ij} \, \frac{\pa}{\pa u^a_j} \ + \ ... \ , $$ are now modified into total derivative operators acting in a larger space,
$$ \wt{D}_i \ = \ D_i \ + \ w^\mu_i \, \frac{\pa}{\pa w^\mu} \ . $$

Note that the equations \eqref{eq:w} have a compatibility
condition; that is, we should require \beql{eq:compH} \wt{D}_i \,
H^\mu_j \ = \ \wt{D}_j \, H^\mu_i \ \ \ \forall \mu = 1,...,r \ ,
\ \ \forall i,j = 1,...,p \ . \eeq The relevant -- interesting and
applicable -- case occurs when these compatibility conditions
\eqref{eq:compH} just amount to the original equations
\eqref{eq:dede}. In this case indeed the original system $\De$ is
properly embedded in the system $\wt{\De}$, or -- seen the other
way round -- $\wt{\De}$ is a \emph{covering} of the original
system $\De$.

\medskip\noindent
{\bf Example 1.} Consider the Gibbons-Tsarev equation \cite{GT} \beql{eq:GTE}  u_{xx} \ + \ u_t \, u_{xt} \ - \ u_x \, u_{tt} \ + \ 1 \ = \ 0 \ ; \eeq
A covering for this is provided by the equations \cite{KraGT,Pryk}
\begin{eqnarray*}
w_t &=& \frac{1}{u_x \ + \ u_t \, w \ - \ w^2} \ := \ H_{(t)} \ ,   \\
w_x &=& \frac{w \ - \ u_t}{u_x \ + \ u_t \, w \ - \ w^2} \ := \ H_{(x)} \ . \end{eqnarray*}
Indeed, if we compute $D_t H_{(x)} - D_x H_{(t)}$ and substitute for $w_t$ and $w_x$ according to the above equations, we obtain
$$ \frac{1 - u_{tt} u_x + u_t u_{xt} + u_{xx}}{[u_x + (u_t - w) w]^2} \ , $$ and immediately recognize that this vanishes if and only if \eqref{eq:GTE} holds. \EOR

\medskip\noindent
{\bf Example 2.} Consider the Burgers equation\footnote{As well
known, this is mapped into the heat equation $v_t = v_{xx}$ by the
Hopf-Cole transformation. Note also that sometimes the equation is
written in a slightly different (potential) form, i.e. as $w_t =
w_{xx} + (1/2) w_x^2$; taking the $x$ derivative of this we get
$w_{xt} = w_{xxx} + w_x w_{xx}$; setting now $u=w_x$ we get
\eqref{eq:Burgers}.} \beql{eq:Burgers} u_t \ = \ u_{xx} \ + \ u \,
u_x \ . \eeq A covering of the Burgers equation is provided by adding the
auxiliary equations written in matrix form as
\beql{eq:exa2b} \frac{\pa W}{\pa x}  \ = \ A \ W \ , \ \ \ \frac{\pa W}{\pa t} \ = \ B \ W \ , \eeq where we have defined the $2 \times 2$ real matrices
\begin{eqnarray*} W &=& \begin{pmatrix} w_{11} & w_{12} \\ w_{21} & w_{22} \end{pmatrix} \ ; \\
A &=& \begin{pmatrix}4 \eta & 2 u + 4\eta \\ 2 u - 4\eta & - 4 \eta \end{pmatrix} \ , \ \
B \ = \ \begin{pmatrix}2 u \eta & u^2 + 2 u_x + 2 u \eta \\ u^2 + 2 u_x - 2 u \eta & - 2 u \eta \end{pmatrix} \ . \end{eqnarray*}
In fact, computing $\chi = D_t [A W] - D_x [B W]$ and then substituting according to \eqref{eq:exa2b}, we immediately obtain that $\chi = 0$ if and only if \eqref{eq:Burgers} holds. \EOR
\bigskip

Coming back to the general discussion, we can now look for standard symmetries of the augmented equation $\wt{\De}$. These will be vector fields to be prolonged in the standard way in the augmented space: thus if $\wt{X}$ is a vector field in $\wt{M} = M \times W = B \times U \times W$, given in local coordinates by
\begin{eqnarray} \wt{X} &=& \xi^i (x,u,w) \, \frac{\pa}{\pa x^i} \ + \ \varphi^a (x,u,w) \, \frac{\pa}{\pa u^a} \ + \ \eta^\mu (x,u,w) \, \frac{\pa}{\pa w^\mu} \nonumber \\ & \equiv & \xi^i \, \pa_i \ + \ \varphi^a \, \pa_a \ + \ \eta^\mu \, \pa_\mu \ , \label{eq:Xtilde} \end{eqnarray}
its prolongation $\wt{Y} \ = \ \wt{X}^{(n)} $ will be a vector field
\begin{eqnarray} \wt{Y} & = & \xi^i \, \frac{\pa}{\pa x^i} \ + \ \psi^a_J \, \frac{\pa}{\pa u^a_J} \ + \ \chi^\mu_J \, \frac{\pa}{\pa w^\mu_J} \nonumber \\ & \equiv & \xi^i \, \pa_i \ + \ \psi^a_J \, \pa_a^J \ + \ \chi^\mu_J \, \pa_\mu^J \ , \label{eq:Ytilde} \end{eqnarray} where $J$ are multi-indices,
$ \psi^a_0 := \varphi^a$, $\chi^\mu_0 := \eta^\mu$, and the coefficients follow the standard prolongation rule, i.e. (recalling the total derivative operators are now the $\wt{D}_i$)
\begin{eqnarray*}
\psi^a_{J,i} &=& \wt{D}_i \psi^a_J \ - \ u^a_{J,k} \, \wt{D}_i \xi^k \ , \\
\chi^\mu_{J,i} &=& \wt{D}_i \chi^\mu_J \ - \ w^\mu_{J,k} \, \wt{D}_i \xi^k \ . \end{eqnarray*}

If such a vector field on $J^n \wt{M}$ is tangent to the solution manifold for the system $\wt{\De}$, i.e. if $\wt{X}$ is a symmetry for $\wt{\De}$, then the restriction of $\wt{X}$ to $M$ will in general be a \emph{nonlocal symmetry} for the equation $\De$ \cite{Kracov,KrV}.

It should be noted that if we just look at the restriction of $\wt{Y}$ to $J^n M$, this is
$$ Y \ = \ \xi^i \, \frac{\pa}{\pa x^i} \ + \ \psi^a_J \, \frac{\pa}{\pa u^a_J} \ \equiv \ \xi^i \, \pa_i \ + \ \psi^a_J \, \pa_a^J \ ; $$
the coefficients $\psi^a_J$ do now appear to follow -- from the point of view of $J^n M$ -- the modified prolongation rule
\begin{eqnarray} \psi^a_{J,i} &=& \( D_i \psi^a_J \ - \ u^a_{J,k} \, D_i \xi^k \) \ + \ w^\mu_i \ \( \pa_\mu \psi^a_J \ - \ u^a_{J,k} \, (\pa_\mu \xi^k ) \) \nonumber \\
&=& \( D_i \psi^a_J \ - \ u^a_{J,k} \, D_i \xi^k \) \ + \ H^\mu_i \ \( \pa_\mu \psi^a_J \ - \ u^a_{J,k} \, (\pa_\mu \xi^k ) \) \ . \label{eq:Ycover} \end{eqnarray}
In the second line, we have used \eqref{eq:w}.

\medskip\noindent
{\bf Remark 2.}
Note that if the $H^\mu_i$ in \eqref{eq:w} are such that their solutions $w^\mu$ can be expressed as a local function of the $u$ -- which in particular is the case if we allow the $H^\mu_i$ to depend also on the $x$-derivatives of the $u$, e.g. $h^\mu_i = c^\mu_a u^a_i$, or if the $H^\mu_i$ depend only on the $x$ but not on the $u$ -- then the above formulas still yield local (albeit not following the standard prolongation formula) prolonged vector fields. \EOR
\bigskip

Finally, we note that one could as well consider \emph{generalized
symmetries}; that is -- with the shorthand notation introduced in
\eqref{eq:Xtilde} -- vector fields
$$ \^X \ = \ \xi^i \, \pa_i \ + \ \varphi^a \, \pa_a \ + \
\eta^\mu \, \pa_\mu $$ where the functions $\xi,\varphi,\eta$
depend not only on $(x,u,w)$ but also on derivatives of $u$ and
$w$ up to some order. If the dependence on derivatives is only in
the $\eta^\mu$, and this is limited to derivatives\footnote{Note
that if the auxiliary equations are first
order, this is automatically true.} of
$u$, i.e.
if we have \beql{eq:Xhat} \^X \ = \ \xi^i (x,u,w) \,
\frac{\pa}{\pa x^i} \ + \ \varphi^a(x,u,w) \, \frac{\pa}{\pa u^a}
\ + \ \eta^\mu (x,u,w;u_x,u_{xx},...) \, \frac{\pa}{\pa w^\mu} \ ,
\eeq then we speak of \emph{semi-classical symmetries}. This will
play a special role in the following, see Section \ref{sec:twistcov} below.

\section{Twisted symmetries}

All different symmetries, Lie-point, non-local, generalized etc.,
considered in the literature share the same fundamental aspect:
there is an action in $M$, and this is lifted -- i.e. prolonged --
to Jet bundles $J^n M$ requiring the prolonged vector field
preserves the contact structure; this requirement is embodied by
the prolongation formula.

It was then rather surprising that in 2001 Muriel and Romero
\cite{MuRom1,MuRom2} proposed a different type of generalization,
where the prolongation formula itself was modified. Starting with
these work (see also \cite{\MuRaltri}), several kinds of
\emph{twisted symmetries} have been considered in the literature
\cite{Gtwist1,Gtwist2}.

For these, one considers a Lie-point vector field $X$ in $M$, but
the prolongation operation is deformed in a way which depends on
some kind of auxiliary object. In different realizations this can
be a scalar function ($\lambda$-symmetries \cite{MuRom1,MuRom2}),
a matrix-valued one-form satisfying the horizontal Maurer-Cartan
equations -- i.e. a set of matrices satisfying a compatibility
condition ($\mu$-symmetries \cite{CGMor}) -- or also a matrix
acting in an auxiliary space ($\sigma$-symmetries
\cite{CGWs1}).\footnote{An actual ``twisting'' only occurs in the
latter cases, not for $\lambda$-symmetries -- where one has
instead a ``stretching'' -- but it is convenient to use this
collective name in all cases where the prolongation operation is
modified \cite{Gtwist1,Gtwist2}.}

It should also be stressed that twisted symmetries are more easily
used for \emph{higher order} differential equations (ordinary or
partial), while the case of first order equations is in some sense
degenerate from this point of view, and presents several
additional problems.

Here we provide a sketchy discussion of different types of twisted
symmetries; the reader can consult e.g. \cite{Gtwist1,Gtwist2} for
further detail and a review.

\subsection{$\la$-symmetries}
\label{sec:lambda}

The first type of twisted symmetries to be introduced was
$\la$-symmetries (the name $C^\infty$ symmetries also appears in
the literature). These were originally introduce to deal with scalar ODEs of any
order, and the name ``$\la$-symmetries'' refers to the auxiliary $C^\infty$ function
$\la (t,x,\xd)$ defining the twisted prolongation, which in this
case is called $\la$-prolongation. In fact, this is recursively
defined as
\begin{eqnarray} \psi^a_{(k+1)} &=& D_x \psi^a_{(k)} \ - \
u^a_{(k+1)} \ D_x \, \xi \ + \ \la \, \( \psi^a_{(k)} \ - \
u^a_{(k)} \, \xi \) \nonumber \\ &=& (D_x \, + \, \la )
\psi^a_{(k)} \ - \ u^a_{(k+1)} \ (D_x \, + \, \la ) \, \xi \ .
\label{eq:laprol}  \end{eqnarray} We will denote the
$\la$-prolongation of order $k$ of the vector field $X$ in $M$ as
$X^{(k)}_{\la}$.

The vector field $X$ in $M$ is said to be a \emph{$\la$-symmetry}
of the equation $\De$ (of order $k$) if \beq X^{(k)}_\la \ : \ S_\De
\ \to \T \, S_\De \ . \eeq Note that in general a vector field is a
$\la$-symmetry of a given equation \emph{only} for a specific
choice of the function $\la$.

\medskip\noindent
{\bf Remark 3.} {In general, the commutator of the
$\la$-prolongations of two vector fields $X,Y$ in $M$ is
\emph{not} the $\la$-prolongation of their commutator, i.e. if $Z
= [X,Y]$ then (in general, for $\la \not= 0$)
\beq \[ X^{(n)}_\la , Y^{(n)}_\la \] \ \not= \ Z^{(n)}_\la \ . \eeq}
In fact, consider e.g. $X = x \pa_u$, $Y = u \pa_u$; in this
case $Z = [X,Y] = X$, and $\delta := [ X^{(1)}_\la ,
Y^{(1)}_\la ] - Z^{(1)}_\la  = x  \la \pa_{u_x} \not= 0$. \EOR
\bigskip

We recall that reduction of ODEs is based on properties of
differential invariants for a prolonged vector field. In
particular, we know that once differential invariants of order
zero and of order one -- call them $\eta$ and $\zeta^{(1)}$ -- are
known, then those of higher orders can be built by just applying
total derivative operators; that is (denoting by $x$ the
independent variable)
$$ \zeta^{(n+1)} \ := \ \frac{D_x \zeta^{(n)}}{D_x \eta} $$
is a differential invariant of order $(n+1)$ if $\zeta^{(n)}$ is a
DI of order $n$ and $\eta$ a DI of order zero. This property,
which stems from the algebra of the prolongation operation, is
also known as ``invariant by differentiation property'', or IBDP.

\medskip\noindent
{\bf Lemma (IBDP Lemma).} {\it The IBDP holds for $\la$-prolonged vector fields.}

\medskip\noindent
{\bf Proof.} This follows from direct computation; see e.g.
\cite{MuRom1,MuRom2}, or \cite{Gtwist1}. \hfill $\diamond$
\bigskip

\medskip\noindent
{\bf Remark 4.} It is the IBDP Lemma that makes $\la$-symmetries ``as useful as
standard ones'', as discussed e.g. in \cite{Gtwist1,Gtwist2}. \EOR

\medskip\noindent
{\bf Remark 5.} It was pointed out by Pucci and Saccomandi
\cite{PuS} that $\la$-prolonged vector fields can be characterized
as the \emph{only} vector fields in $J^k M$ with the property that
their integral lines are the same as the integral lines of some
vector field which is the standard prolongation of some vector
field in $M$. This remark was fully understood only some time
after their paper, and was the basis for many of the following
developments, discussed below. \EOR

\subsection{$\mu$-symmetries}
\label{sec:mu}

The $\la$-prolongation is specifically designed to deal with ODEs
(or systems thereof); a generalization of it aiming at tackling
PDEs (or systems thereof) is the $\mu$-prolongation. This can of
course also be applied to ODEs and Dynamical Systems, as we will
see below.

\subsubsection{PDEs}

Now the relevant object is not a single matrix, but an array of
matrices $\La_i$, one for each independent variable. These are
better encoded as a ($GL(n,\R)$-valued) \emph{horizontal one-form}
\beql{eq:mu} \mu \ = \ \La_i (x,u,u_x) \, \d x^i \ . \eeq The
matrices $\La_i$ should satisfy a compatibility condition, i.e.
\beql{eq:MCH} D_i \, \La_j \ - \ D_j \, \La_i \ + \ \left[
\Lambda_i , \Lambda_j \right] \ = \ 0 \ ; \eeq this is immediately
recognized as the \emph{horizontal Maurer-Cartan equation}, or
equivalently as a \emph{zero-curvature condition} for the
connection on $\T U$ identified by \beql{eq:nabla} \nabla_i \ = \
D_i \ + \ \Lambda_i \ . \eeq

If $\mu$ satisfies \eqref{eq:MCH}, we can define
$\mu$-prolongations in terms of a modified prolongation formula,
called of course \emph{$\mu$-prolongation formula} (and which
represents now an actual twisting of the familiar prolongation
operation): \begin{eqnarray} \psi^a_{J,i} &=& D_i \psi^a_{J} \ -
\ u^a_{J,k} \ D_i \, \xi^k \ + \ (\La_i)^a_{\ b} \, \( \psi^b_{J}
\ - \ u^b_{J,k} \, \xi^k \) \nonumber \\ &=& (D_i \, I \, + \, \La_i )^a_{\
b} \, \psi^b_{J} \ - \ u^b_{J,k} \ (D_i \, I \, + \, \La_i )^a_{\
b} \, \xi^k \ . \label{eq:muprol} \end{eqnarray}

We will denote the $\mu$ prolongation (of order $k$) of the vector
field $X$ in $M$ as $X^{(k)}_\mu$. The vector field $X$ in $M$ is
said to be a \emph{$\mu$-symmetry} of the equation $\De$ (of order
$k$) if \beq X^{(k)}_\mu \ : \ S_\De \ \to \T \, S_\De \ . \eeq Note
that in general a vector field is a $\mu$-symmetry of a given
equation \emph{only} for a specific choice of the one-form $\mu$.

\medskip\noindent
{\bf Remark 6.} In $\la$-prolongations the prolongation operation
is modified, but it acts separately on the different vectorial
components in $\T U$ (and in $\T U_J$). In $\mu$-prolongations,
instead, the different vector components of $\T U$ (and of $\T
U_J$) are ``mixed'' by the prolongation operation which thus
operates a ``twisting'' among different components of the vector
field; this is the origin of the name ``twisted symmetries''.
Obviously, $\la$-symmetries are -- even in the vector framework --
a special case of $\mu$-symmetries, with matrices $\La_i$ being
multiple of the identity matrix through functions $\la_i$. \EOR

\medskip\noindent
{\bf Remark 7.} It is known that $\mu$-symmetries (and hence
$\la$-symmetries) are related to \emph{nonlocal} symmetries
\cite{CF,MuRom5,MuRom12}; we will discuss this relation below.
\EOR

\subsubsection{ODEs}

In the case of ODEs one just replaces the scalar function $\la :
J^1 M \to \R$ with a \emph{matrix} function $\La : J^1 M \to
\mathtt{Mat} (n)$ and define a ``$\La$-prolongation'' \cite{Cds1,Cds2}
(which is just a special case of
$\mu$-prolongation, for $\mu = \La \d x$)
\begin{eqnarray} \psi^a_{(k+1)} &=& D_x
\psi^a_{(k)} \ - \ u^a_{(k+1)} \ D_x \, \xi \ + \ \La^a_{\ b} \,
\( \psi^b_{(k)} \ - \ u^b_{(k)} \, \xi \) \nonumber \\ &=& (D_x \, I \, + \,
\La )^a_{\ b} \, \psi^b_{(k)} \ - \ u^b_{(k+1)} \ (D_x \, I \, +
\, \La )^a_{\ b} \, \xi \ . \label{eq:Laprol} \end{eqnarray}

In this ODE case we just have $\mu = \La \, \d x $
(only one component), and \eqref{eq:MCH} is identically satisfied.

\medskip\noindent
{\bf Remark 8.} The IBDP property is in general \emph{not} holding
for $\mu$-prolonged vector fields, not even in the ODEs framework;
the exception is the case where the $\La_i$ are diagonal matrices.
This means that in general $\mu$-symmetries can not be used to
obtain a symmetry reduction of ODEs (see however Remark 9 below).
\EOR

\subsubsection{Recursion formula}

The $\mu$-prolongation $X^{(k)}_\mu$, which we will now write in
components as $X^{(k)}_\mu = \xi^i \pa_i + (\psi^a_J)_{(\mu)}
\pa_a^J $, of a vector field $X$ in $M$ is defined through
\eqref{eq:muprol}; however in some cases and applications it is
relevant to characterize these in terms of the difference \beq
F^a_J \ := \ \( \psi^a_J \)_\mu \ - \ \( \psi^a_J \)_0 \ . \eeq It
can be shown \cite{CGMor,GMor} that the $F^a_J$ satisfy the
recursion relation \beql{eq:Fmu} F^a_{J,i} \ = \ \de^a_b \ \[ D_i
\, \( \Ga^J \)^b_c \] \ (D_i Q^c )  \ + \ \( \La_i \)^a_b \ \[
\(\Ga^J \)^b_c \ (D_J Q^c ) \ + \ D_j Q^b \]   \ , \eeq where we
have written \beql{eq:Q} Q^a \ = \ \vphi^a \ - \ u^a_i \, \xi^i \
, \eeq and the $\Ga^J$ are certain matrices whose detailed
expression can be computed \cite{CGMor,GMor} but is not essential.

\medskip\noindent
{\bf Remark 9.} With the notation \eqref{eq:Q}, the set $I_X$ of
$X$-invariant functions is characterized by $Q^a |_{I_X} = 0$. It
follows from \eqref{eq:Fmu} that $X^{(k)}_\mu$ coincides with
$X^{(k)}_0$ on $I_X$. This means that $\mu$-symmetries are as good
as standard Lie-point symmetries to obtain invariant solutions to
differential equations -- which is what we do when we have determined
symmetries of PDEs. \EOR

\subsection{$\s$-symmetries}

When dealing with symmetries of differential equations we often
use them one at a time, in particular for ODEs -- e.g. when we
reduce the order of the equation. But in general we have a
$k$-dimensional Lie algebra $\G$ of symmetries; the prolongation
acts separately on each vector field in $\G$.

It turns out that a different kind of modification of the
prolongation operation is possible when we consider a Lie algebra
$\G$ of vector field, or more generally a system of vector fields
which are in involution (in the sense of Frobenius); in this case
the ``twisting'' corresponds to mixing the different vector fields
in the prolongation operation. This approach has received the name
of ``$\s$-prolongation'' and correspondingly one speaks of
``$\s$-symmetries'' \cite{CGWs1,CGWs2,CGWs3,CGWs4}. This approach
is specially suited to the study of dynamical systems.

We will not discuss this type of twisted symmetries here; the
reader is referred to the original papers cited above or to the reviews
\cite{Gtwist1,Gtwist2}.

\section{Twisted prolongations and gauge groups}

Let us consider the case where the fields $u^a = u^a(x)$, i.e. the
dependent variables, take values in a vector space $U = R^q$; in
this case $M$ is a vector bundle.\footnote{The general case can be
treated along the same lines; but as our considerations will be
local, this would just lead to a heavier notation and discussion.}

We can then operate an $x$-dependent change of frame in $U$; as
well known, this means acting on our fields (and equations) by a
\emph{gauge transformation}.

When we deal with $J^n M$, there are natural coordinates $u^a_J$
in it. Note that for a given multi-index $J$ the variables $u_J = (u^1_J,...,u^q_J)$
can be seen as belonging to a vector space $U_J$ isomorphic to $U$; we can then
prolong the gauge transformation defined on $U$ (more precisely,
on the bundle $(M,\pi,B)$) to a gauge transformation in $J^N M$
(more precisely, on the bundle $(J^n M, \pi_n , B)$) by acting in
the same way on all the vector spaces $U_J$, $|J| = 0, ... , n$.

This induces an action on vector fields on $M$ as well as on
vector fields on $J^n M$; it is rather obvious that such an action
is specially simple if we look at \emph{vertical} vector fields,
including the evolutionary representative
$$ X_v \ = \ (\varphi^a \ - \ u^a_i \xi^i) \ \frac{\pa}{\pa u^a}
\ := \ \phi^a (x,u,u_x) \ \frac{\pa}{\pa u^a} $$ of any Lie-point
vector field $$ X \ = \ \xi^i (x,u) \, \frac{\pa}{\pa x^i} \ + \
\varphi^a (x,u) \, \frac{\pa}{\pa u^a} $$ in $M$.\footnote{Note
that, as well known, $X_v$ is in general (that is, unless $\xi^i =
0$ for all $i = \,...,p$) a generalized vector field, and the
formalism of evolutionary representatives has full geometrical
sense only when considering infinite jets $J^\infty M$
\cite{KrV}.}

Let us thus consider vector fields $X$ on $M$ and their
prolongations $X^{(n)}$  on $J^n M$, or better the evolutionary
representatives $X_v$ and their prolongations $X_v^{(n)}$; and let
us consider the gauge transformed of these. Due to the
\emph{local} nature of the gauge transformation, the gauge
transformed of $X_v^{(n)}$ is \emph{not} the prolongation of the
gauge transformed of $X_v$.

Let us denote the $\mu$-prolongation operator defined in Section
\ref{sec:mu} as $\mathtt{Pr}_\mu$, with $\mathtt{Pr} =
\mathtt{Pr}_0$ the standard prolongation operator, and denote by
$\ga$ a given gauge transformation.

Then it turns out that the diagram (where now all vector fields
are vertical, albeit this is not explicitly indicated in order to
keep notation simple) \beql{eq:diagram} \begin{matrix}
X & \mapright{\ga} & W \\
\mapdown{{\tt Pr}_0} & & \mapdown{{\tt Pr}_\mu} \\
Y & \mapright{\ga} & Z \end{matrix}
 \eeq
is commutative, provided $\ga = R^a_{\ b} (x,u)$ and $\mu$ are related by
\beql{eq:gaugemu} \mu \ = \ R^a_{\ c} \[  D_i \, (R^{-1})^c_{\ b} \] \
\d x^i \ := \ \La_i \ \d x^i \ . \eeq

This is readily seen for first prolongations\footnote{And hence
for higher ones as well, recalling that the $(n+1)$-th
prolongation is the first prolongation of the $n$-th
prolongation.} just working in coordinates. We write
$$ X \ = \ \phi^a \, \frac{\pa}{\pa u^a} \ , \ \ \ W \ = \
(R^a_{\ b} \,\phi^b ) \, \frac{\pa}{\pa u^a} \ ; $$
the (standard) first prolongations of these are respectively
\begin{eqnarray*}
Y \ = \ X^{(1)} &=& \phi^a \, \frac{\pa}{\pa u^a} \ + \ (D_i \phi^a) \, \frac{\pa}{\pa u^a_i} \ , \\
Z \ = \ W^{(1)} &=& (R^a_{\ b} \phi^b ) \, \frac{\pa}{\pa u^a} \ + \ [D_i (R^a_{\ b} \phi^b) ] \, \frac{\pa}{\pa u^a_i} \\
&=& R^a_{\ b} \phi^b \, \frac{\pa}{\pa u^a} \ + \ R^a_{\ b} (D_i \phi^b) \, \frac{\pa}{\pa u^a_i} \ + \ (D_i R^a_{\ b} ) \, \phi^b \, \frac{\pa}{\pa u^a_i} \\
&=& R^a_{\ b} \[ \phi^b \, \frac{\pa}{\pa u^a} \ + \ (D_i \phi^b) \, \frac{\pa}{\pa u^a_i} \] \ + \ \[ (D_i R^a_{\ \ell} ) \, (R^{-1})^\ell_{\ m} \, R^m_{\ b} \, \phi^b \] \, \frac{\pa}{\pa u^a_i} \ . \end{eqnarray*}
On the other hand, it is immediate to see that the gauge transformed of $Y$ is
$$ \ga (Y) \ = \ R^a_{\ b} \[ \phi^b \, \frac{\pa}{\pa u^a} \ + \ (D_i \phi^b) \, \frac{\pa}{\pa u^a_i} \] \ ; $$ thus in order to have a commutative diagram we need to choose
$$ \mu \ = \ - \, (D_i R) \, R^{-1} \, \d x^i \ = \  R \, D_i R^{-1} \, \d x^i \ . $$
In other words, the matrices $\La_i$ in the definition of the
horizontal one-form $\mu$ must be chosen according to
\eqref{eq:gaugemu}.

As mentioned above, this computation extends at once to higher
order prolongations.

\medskip\noindent
{\bf Remark 10.} Note that the compatibility condition discussed
in Section \ref{sec:mu} is automatically satisfied. In fact, now
\begin{eqnarray*}
D_i \La_j \ - \ D_j \La_i &=& D_i \, (R \, D_j R^{-1}) \ - \
D_j \, (R \, D_i R^{-1} ) \\
&=& (D_i R) \, (D_j R^{-1}) \ + \ R \, (D_i D_j R^{-1}) \ - \
(D_j R) \, (D_i R^{-1}) \ - \ R \, (D_j D_i R^{-1} ) \\
&=& (D_i R) \, (D_j R^{-1}) \ - \ (D_j R ) \, (D_i R^{-1} ) \ ; \\
\[ \La_i , \La_j \] &=& R \, (D_i R^{-1} ) \cdot R \, (D_j R^{-1} )
\ - \ R \, (D_j R^{-1} ) \cdot R \, (D_i R^{-1} ) \\
&=& - \, R [R^{-1} \,(D_i R) \, R^{-1} ] \, R \, (D_j R^{-1}) \ + \
R [R^{-1} \, (D_j R) \, R^{-1} ] \, R \, (D_i R^{-1}) \\
&=& - \, (D_i R) \, (D_j R^{-1}) \ + \
(D_j R) \, (D_i R^{-1}) \ . \end{eqnarray*}
Thus the horizontal Maurer-Cartan equation \eqref{eq:MCH} is satisfied. \EOR
\bigskip

We summarize our discussion in the form of the following statements (their proof is in fact given by the previous discussion):

\medskip\noindent
{\bf Proposition 1.} {\it $Z$ be the $\mu$-prolongation of the vertical vector field $W$, defined on $(M,\pi,B)$, to $J^n M$. Then there are vertical vector fields $X$ on $M$ and $Y$ on $J^n M$ which are gauge-equivalent to $W$ and $Z$ respectively, and such that $Y$ is the standard prolongation of $X$. The gauge transformation realizing this equivalence and the horizontal one-form $\mu$ in $J^1 M$ are related by \eqref{eq:gaugemu}.}

\medskip\noindent
{\bf Corollary.} {\it Let $W$ be a $\mu$-symmetry for a given differential equation $\De$. Then there is a vector field $X$ on $M$ such that a gauge transform of its standard prolongation is tangent to $S_\De \ss J^n M$.}

\section{Twisted prolongations and gauging of \\ derivatives}

A different approach, also based on gauge transformations, has
been followed by Morando \cite{Mordef}. She noted that one can
describe $\la$ and $\mu$ symmetries in terms of gauge-deformed Lie
and exterior derivatives. We will follow her work, and work
directly with $\mu$-prolongations and $\mu$-symmetries; as already
mentioned, this includes $\la$-prolongations and $\la$-symmetries
as a special case.

In the case of $\mu$-prolongations, the fundamental object is the
closed differential horizontal one-form $\mu = \La_i \d x^i$. One
can define a deformed exterior derivative $\d_\mu$ acting on forms
of any degree by \beq \d_\mu \a \ := \ \d \a \ + \ \mu \w \a \ .
\eeq It is immediate to check that $\d_\mu^2 = 0$; thus $\d_\mu$
allows to define a cohomology.

When $\mu = \d f$, with $f$ a $\C^\infty$ function on $B$, we have
$$ \d_\mu \a \ = \ e^{-f} \ \d ( e^f \a) \ ; $$
in this sense the deformed exterior derivative $\d_\mu$
corresponds to (a generalization of) a gauging of the standard
exterior derivative $\d$.

Similarly, one can consider a deformed Lie derivative $\L^\mu$. For
$X$ a vector field, the deformed Lie derivative $\L_{X}^\mu$ is
defined to act on forms $\a$ and on vector fields $Y$ by
\begin{eqnarray*} \L_{X}^\mu ( \a ) & = & \L_{X} \a \ + \ \mu \w (X \interno \a) \ , \\
\L_{X}^\mu (Y) & = & \L_{X} (Y) \ - \ (Y \interno \mu) \ X \ . \end{eqnarray*}
Again, if $\mu = \d f$ these read
\begin{eqnarray*} \L_{X}^\mu ( \a ) & = & e^{- f} \ \L_{(e^f X)} \ ( \a ) \ , \\
\L_{X}^\mu (Y) & = & e^{- f} \ \L_{(e^f X)} (Y)  \ , \end{eqnarray*}
so this corresponds to (a generalization of) a gauging of the standard Lie derivative $\L$.

Then, $\mu$-prolonged vector fields can be characterized exactly
in the same way as standardly prolonged ones, at the exception
that the deformed Lie derivative plays the role of the standard
one.

That is, we consider the contact forms $\vartheta^a_J = \d u^a_J
-u^a_{J,i} \d x^i $ and the Cartan ideal $\Theta$ generated by
them. The we have:

\medskip\noindent
{\bf Proposition 2.} {\it A vector field $Y$ on $J^n M$ is the $\mu$-prolongation of
the vector field $X$ in $M$ if and only if \par\noindent $(a)$ it admits a
projection on $M$, and this coincides with $X$; \par\noindent $(b)$ it satisfies
$$ \L_Y^\mu (\Theta ) \ \sse \ \Theta \ , $$
i.e. for any $a,J$  there are smooth functions $A^{\mu K}_{J,\b}$ such that
$$ \L_Y^\mu (\vartheta^a_J ) \ = \ A^{\mu K}_{J \b} \ \vartheta^\b_K \ . $$}

\medskip\noindent
{\bf Proof.}
This is Theorem 4 in \cite{Mordef}, and the reader is referred to
there for a proof, extensions, and a discussion. \hfill $\diamond$

\section{Twisted prolongations and coverings}
\label{sec:twistcov}

The theory of coverings allows to provide a nonlocal
interpretation of $\la$ and more generally $\mu$ symmetries; that
is, a (local) $\mu$ symmetry for a given equation corresponds to a
standard \emph{non-local} one for the same equation. This
generalizes a property holding also for standard symmetries
\cite{Kracov,KrV}.

The idea is the following. If the auxiliary equations
\eqref{eq:w} are solved for $w$ as a function of the $x$ and $u$,
say with \beql{eq:wsol} w^\mu \ = \ \Theta^\mu (x,u) \ , \eeq then
we can restrict the vector field $\wt{X}$, see \eqref{eq:Xtilde},
to the $(x,u)$ space; this will be \beq \wt{X}_0 \ = \ \xi^i \[ x
, u , \Theta (x,u) \] \, \frac{\pa}{\pa x^i} \ + \ \varphi^a \[ x
, u , \Theta (x,u) \] \, \frac{\pa }{\pa u^a} \ . \eeq

But in general -- albeit not always -- the functions $\Theta^\mu
(x,u) $ will contain \emph{integrals} of $x$ and $u$, as some
trivial or less trivial example can easily show.

\medskip\noindent
{\bf Example 3.} Consider the equation
\beql{eq:ex3a} du/dx \ = \ f (x,u) \ = \ u \ ; \eeq
we add to this the equation
\beql{eq:ex3b} d w / dx \ = \ h (x,u,w) \ = \ u \, w \ ; \eeq
note that the latter is rewritten as
$ dw/w = u d x $ and hence solved by \beql{eq:ex3c} w(x) \ = \ \exp \[ \int u \, d x \] \ .
\eeq Consider now Lie-point symmetries for the system
\eqref{eq:ex3a}, \eqref{eq:ex3b}; these will be in the form
\eqref{eq:Xtilde}. One of the symmetries of the system turns out
to be\footnote{The action of this vector field is
readily integrated to give $w(s) = k_1 e^s$, $u(s) = k_2
\exp[w[s]]$; the quantity $u e^{-w}$ is thus invariant under
$\wt{X}$.}
$$ \wt{X} \ = \ u \, w \, \pa_u \ + \ w \, \pa_w \ ; $$
by using \eqref{eq:ex3c}, the restriction of this to the $(x,u)$ space is
\beq \wt{X}_0 \ = \ \( u \, \exp \[ \int u \, dx \] \) \ \pa_u \ , \eeq
i.e. a non-local vector field. \EOR

\medskip\noindent
{\bf Example 4.} (See \cite{KrV}, Section 6.1.) Let us consider
again the Burgers equation
$$ u_t \ = \ u_{xx} \ + \ u \, u_x \ . $$
Then we have symmetries
$$ X_\a \ := \ \( \a \, u \ - \ 2 \, \a_x \) \ \exp \[ - \frac12 \, \int u dx \] \
\frac{\pa}{\pa u} \ , $$
with $\a = \a (x,t)$ any solution to the heat equation $\a_t = \a_{xx}$.

If we look for solutions to the Burgers equation which are
invariant under $X_\a$, we have to solve for the system made of
the Burgers equation and of the condition $X_\a [u] = 0$, i.e.
\begin{eqnarray*}
& & u_t \ = \ u_{xx} \ + \ u \, u_x  \\
& & \( \a \, u \ - \ 2 \, \a_x \) \ \exp \[ - \frac12 \int u d x \] \ = \ 0 \ . \end{eqnarray*}
The second equation requires
$u  =  2 \a_x / \a $;
plugging this into the first one, we obtain
$$ \frac{2}{\a^2} \ \[ \a \, D_x \( \a_t \, - \, \a_{xx} \) \ - \ \a_x \ \( \a_t \, - \, \a_{xx} \) \]
\ = \ 2 \ D_x \( \frac{\a_t \, - \, \a_{xx}}{\a} \) \ . $$ In
other words, the nonlocal symmetries $X_\a$ lead us to the Hopf-Cole transformation. \EOR
\bigskip

\subsection{$\la$-symmetries}

Pretty much the same mechanism is at work also when one considers
twisted rather than standard symmetries. In particular the
situation is fully understood in the case of $\la$-symmetries
(while no much work in the context of $\mu$ and
$\s$-symmetries appears in the literature, see however the next subsection);
in this context we have the following general
result, which is Proposition 1 in \cite{CF}.

\medskip\noindent
{\bf Proposition 3.} {\it Consider a given smooth function $\la =
\la (x,u,u_x)$; consider moreover the ODE
$$ \De_0 \ := \ \ \ \ \frac{d^k u}{d x^k} \ = \ f(x,u,...,u^{(k-1)} ) $$
and its covering $\wt{\De}$ consisting of the system
\begin{eqnarray*} \frac{d^k u}{d x^k} &=& f(x,u,...,u^{(k-1)}
)\\ \frac{d w}{d x} &=& \la (x,u,u_x ) \ . \end{eqnarray*}
Then $\De$
admits a $\la$-symmetry $X$ if and only if $\wt{\De}$ admits a
\emph{semi-classical} symmetry $Y = \xi \pa_x + \varphi \pa_u +
\eta \pa_w$ such that $[\pa_w , Y] = Y$. Moreover, $X$ is the
projection to the $(x,u)$ space of the restriction of $Y$ to the
solution manifold for the auxiliary equation $dw/dx =
\la(x,u,u_x)$, i.e. to $$ w (x) \ = \ \int \la (x,u,u_x) \, d x \
. $$}

\medskip\noindent
{\bf Proof.} For a detailed proof, the reader is referred to \cite{CF}. Here we give a sketch of it.
For a given equation $\De_0$, we consider the system $\wt{\De}$ consisting of it and of $\De_1$ given by $w_x = \la (x,u,u_x)$. Suppose then that some Lie-point symmetry $X = \xi (x,u,w) \pa_x + \varphi (x,u,w) \pa_u + \eta (x,u,w) \pa_w$ for $\wt{\De}$ has been determined, and denote by $Y$ the prolongation (of suitable order) of $X$. This means that
$$ \[ Y (\De_0 )\]_{\{ \De_0 = 0 , \De_1 = 0\}} \ = \ 0 \ , \ \ \ \[ Y (\De_1 )\]_{\{ \De_0 = 0 , \De_1 = 0\}} \ = \ 0 \ . $$
On the other hand, it is clear that $Y (\De_0 )$ only involves the prolongation of $X_0 = \xi (x,u,w) \pa_x + \varphi (x,u,w) \pa_u$, call it $Y^{(0)}$. This is of the form
$$ Y^{(0)} \ = \ \xi \, \pa_x \ + \ \sum_k \psi^{(k)} \, \frac{\pa }{\pa u^{(k)}} \ , $$
where $\psi^{(0)} = \varphi$ and the $\psi^{(k)}$ obey the prolongation formula
\beql{eq:prolw} \psi^{(k+1)} \ = \ D_x \psi^{(k)} \ - \ u^{(k+1)} \ D_x \xi \ . \eeq
It is convenient to rewrite the total derivative operator
$$ D_x \ = \ \pa_x \ + \ \sum_k u^{(k+1)} \, \frac{\pa }{\pa u^{(k)}} \ + \ \sum_k w^{(k+1)} \, \frac{\pa}{\pa w^{(k)}} $$ in the form
\beql{eq:D0D1a} D_x \ = \ D_x^{(0)} \ + \ D_x^{(1)} \ , \eeq having defined
\beql{eq:D0D1b} D_x^{(0)} \ = \ \pa_x \ + \ \sum_k u^{(k+1)} \, \frac{\pa }{\pa u^{(k)}} \ ; \ \ \ D_x^{(1)} \ = \ \sum_k w^{(k+1)} \, \frac{\pa}{\pa w^{(k)}} \ . \eeq
With this notation, we rewrite eq.\eqref{eq:prolw} as
\beql{eq:prolw2} \psi^{(k+1)} \ = \ D_x^{(0)} \psi^{(k)} \ - \ u^{(k+1)} \ D_x^ {(0)} \xi \ + \
D_x^{(1)} \psi^{(k)} \ - \ u^{(k+1)} \ D_x^{(1)} \xi \ . \eeq

If we assume that the condition
$ \[ Y (\De_1 )\]_{\{ \De_0 = 0 , \De_1 = 0\}} =  0 $ is satisfied, the other condition
$ \[ Y (\De_0 )\]_{\{ \De_0 = 0 , \De_1 = 0\}} =  0 $ can be rewritten solving explicitly $\De_1$ as
$$ \[ Y (\De_0 )\]_{\{ \De_0 = 0 , w = \int \la d x \}} \ = \ 0 \ . $$ This in turn can be written as
$$ \[ \^Y (\De_0 )\]_{\{ \De_0 = 0 \}} \ = \ 0 \ , $$
where the vector field $\^Y$ is defined by restricting the vector field $Y$ to
\beql{eq:wsol} w \ = \ \int \la (x,u,u_x) \, dx \eeq and its differential consequences; note that under this restriction we get
\beq D_x^{(1)} \ = \ \la \, \pa_w \ + \ (D_x \la) \, \pa_{w_x} \ + \ ... \ = \ \sum_\ell (D_x^\ell \la ) \, \frac{\pa}{\pa w^{(\ell)} } \ . \eeq Thus \emph{if} $[\pa_w , Y ] = Y$, it follows that $\varphi$ and $\xi$ are of the form
\beql{eq:expw} \varphi (x,u,w) \ = \ e^w \ \varphi_0 (x,u) \ , \ \ \ \xi (x,u,w) \ = \ e^w \ \xi_0 (x,u) \ , \eeq and then \eqref{eq:prolw2} reads just as the $\la$-prolongation formula. \footnote{Note that the same condition $[\pa_w , Y] = Y$ also implies $\eta (x,u,w) = e^w \eta_0 (x,u)$.} \hfill $\diamond$
\bigskip

The situation can be summarized in a diagram:
$$ \begin{matrix} \wt{\De} & \mapright{{\tt sym}} & \wt{X} & \mapright{{\tt Pr}_0} & \wt{Y} \\
\mapdown{{\tt cov}} & & \mapdown{\De_1 = 0} & & \mapdown{\De_1 = 0} \\
\De_0 & \mapright{\la-{\tt sym}} & X^{(0)} & \mapright{{\tt Pr}_\la} & Y^{(0)} \end{matrix} $$
Here ${\tt sym}$ (respectively, $\la - {\tt sym}$) refers to the fact we determine a symmetry (a $\la$-symmetry) of the equation, ${\tt cov}$ refers to the fact $\wt{\De}$ is a covering of $\De_0$, and $\De_1=0$ refers to the restriction to the solution manifold for $\De_1$ (and its differential consequences). Note here $\wt{X}$ must be of the form \eqref{eq:expw}.

We will illustrate this result by an example, also taken
from \cite{CF}, which we consider in some detail.

\medskip\noindent
{\bf Example 5.} Consider the equation, or actually the class of equations,
\beql{eq:AGL} \De := \ \ \ u_{xx} \ = \ \frac{u_x^2}{u} \ + \ \[ m \, g(x) \, u_x \ + \ g'(x) \, u \] \, u^m \ , \eeq
where $g(x)$ is a smooth function and $m \not= 0$ a real constant. This class of equations was studied by Gonzalez-Lopez \cite{GL}, and for general $g(x)$ it has no Lie-point symmetries. On the other hand, it was shown by Muriel and Romero \cite{MuRom1}, and it is easily checked, that it always admits as $\la$-symmetry the vector field
$$ X \ = \ \pa_u $$ provided one chooses
$$ \la (x,u,u_x) \ = \ \frac{u_x}{u} \ + \ m \, g(x) \, u^m \ . $$

In fact, the second $\la$-prolongation of $X$ will be
$$ Y \ = \ \pa_u \ + \ \^\psi^{(1)} \, \pa_{u_x} \ + \ \^\psi^{(2)} \, \pa_{u_{xx}} \ , $$ with the coefficients $\psi^{(k)}$ satisfying the $\la$-prolongation formula, which in this case ($\xi = 0$) reads simply
$$ \psi^{(k+1)} \ = \ D_x \psi^{(k)} \ + \ \la \, \psi^{(k)} \ , $$ and of course with $\psi^{(0)} = 1$. Thus we get
$$ \psi^{(1)} \ = \ \la \ , \ \ \psi^{(2)} \ = \ D_x \la \ + \ \la^2 \ . $$
Thus, by explicit computation,
$$ Y[\De] \ = \ \frac{u \, u_{xx} \ - \ u_x^2 \ - \ u^{m+1} \[ m \, g(x) \, u_x \ + \ u \, g' (x) \]}{u^2} \ ; $$ substituting for $u_{xx}$ according to $\De$ -- i.e. according to eq.\eqref{eq:AGL} -- we get indeed
$$ \[ Y [\De] \]_{\De = 0} \ = \ 0 \ . $$

When we consider the system $\wt{\De}$ made by \eqref{eq:AGL} and by the auxiliary equation
\beql{eq:AGLb} w_x \ = \ \la (x,u,u_x ) \eeq and look for standard Lie-point symmetries, say of the simplified form
$$ \wt{X} \ = \ \varphi (x,u,w) \, \pa_u \ + \ \eta (x,u,w) \, \pa_w $$
it turns out that choosing
$$ \varphi \ = \ e^w \ , \ \ \eta \ = \ (m+1) \ \frac{e^w}{u} \ , $$
or in other words
$$ \wt{X} \ = \ e^w \ \[ \pa_u \ + \ \frac{m+1}{u} \, \pa_w \] \ , $$ we have a symmetry. This can be checked by standard computations.

On the other hand, \eqref{eq:AGLb} is solved by
\beql{eq:AGLw} w \ = \ \int \la (x,u,u_x) \, dx \ = \ \log (u) \ + \ m \, \int u(x) \, g(x) \, d x \ ; \eeq thus the vector field $\wt{X}$ restricted to the solution to \eqref{eq:AGLb} and projected to the $(x,u)$ space reads
$$ \^X \ = \ \exp \[ \la \, d x \] \ \pa_u \ , $$
i.e. we have a non-local vector field.

Now if we look at the second prolongation of $\wt{X}$, we have
\begin{eqnarray*} \wt{Y} &=& e^w \ \[ \frac{\pa}{\pa u} \ + \ w_x \, \frac{\pa}{\pa u_x} \ + \ (w_x^2 + w_{xx} )  \, \frac{\pa}{\pa u_{xx}} \] \\ & & + \ e^w \ \frac{(m+1)}{u} \ \[ \frac{\pa}{\pa w} \ + \ \frac{u w_x - u_x}{u} \, \, \frac{\pa}{\pa w_x} \right. \\
& & \left.  + \ \frac{2 u_x^2 - 2 u u_x w_x - u u_{xx} + u^2 (w_x^2 + w_{xx})}{u^2} \, \frac{\pa}{\pa w_{xx}} \] \ . \end{eqnarray*}

When we restrict to solutions to \eqref{eq:AGLb}, i.e. substitute for $w$ and its derivatives according to \eqref{eq:AGLw}, and project to the $(x,u,u_x,u_{xx})$ space -- i.e. to $J^2 M$ -- we get
\beq \wt{Y} \ = \ \( \exp \[ \int \la dx \] \) \ \[ \frac{\pa}{\pa u} \ + \ \la \, \frac{\pa}{\pa u_x} \ + \ (\la^2 + D_x \la )  \, \frac{\pa}{\pa u_{xx} } \] \ . \eeq
By construction, this is tangent to the solution manifold for $\De$, $\wt{Y} : S_\De \to \T S_\De$. But if this is the case, the same also applies to any vector field which is collinear to $\wt{Y}$, in particular to
\begin{eqnarray} \^Y &=& \exp \[ - \int \la \, dx \] \ \^Y \nonumber \\
&=& \frac{\pa}{\pa u} \ + \ \la \, \frac{\pa}{\pa u_x} \ + \ (\la^2 + D_x \la )  \, \frac{\pa}{\pa u_{xx} } \ . \end{eqnarray}
This is the $\la$-prolongation of the vector field $\^X = \pa_u$. \EOR
\bigskip

\subsection{$\mu$-symmetries}

The discussion given above for $\la$-symmetries can be extended to $\mu$-symmetries,
provided we only consider \emph{vertical} vector fields, both in the $(x,u)$ space and in the augmented $(x,u,w)$ one.

Thus to a PDE or system of PDEs $\De_0$ of order $n$
\beql{eq:D0mumu} \De_0 \ : \ \ \ \ F^\ell (x,u,...,u^{(n)} ) \ = \ 0 \ , \ \ \ \ \ell = 1 ,..., L \eeq
for $u = (u^1 , ... , u^p ) $ depending on the independent variables $x = (x^1,...,x^q)$ we associate the auxiliary equations for $w = (w^1 , ... , w^m)$ given by
\beql{eq:D1mumu} \De^\beta_i \ : \ \ \ \ w^\beta_i \ = \ h^\beta_i (x,u,w,u_x) \ , \eeq
where the functions $h^\beta_i$ satisfy the compatibility condition
\beql{eq:compmumu} D_i h^\beta_j \ = \ D_j h^\beta_i \eeq for all pairs $i,j=1,...,q$ and for all $\mu = 1,...,m$. Note that now and in the following $D_i$ denotes the total derivative w.r.t. $x^i$ in the augmented space, i.e. taking care of both the $u$ and the $w$ variables.

We will then consider the system $\wt{\De}$ made of the original equation $\De_0$ and of the auxiliary equations $\De^\beta_i$. When looking for Lie-point symmetries of $\wt{\De}$, we will \emph{only} be considering vertical vector fields, i.e. vector fields of the form
\beql{eq:Xmumu} X \ = \ \phi^a (x,u,w) \, \frac{\pa}{\pa u^a} \ + \ \eta^\beta (x,u,w) \, \frac{\pa}{\pa w^\beta} \ . \eeq
In order to apply this to $\wt{\De}$, it suffice to consider prolongation to order $n$ in the $u$ derivatives but only to order one in the $w$ derivatives; we will write this as
\beq Y \ = \ \Psi^a_J \, \frac{\pa}{\pa u^a_J} \ + \ \chi^\beta_i \, \frac{\pa}{\pa w^\beta_i} \ , \eeq
where $J$ is a multi-index of order $|J| \le n$, the index $i$ runs on $1,...,q$, and sum over repeated indices and multi-indices is understood. Moreover we set $\Psi^a_0 = \Phi^a$, $\chi^\beta_0 = \eta^\beta$.  We will also write, for later reference, the restriction of $Y$ to the $J^n M$ bundle (with $M = B \times U$, and $x \in B$, $u \in U$ the manifolds in which $x$ and $u$ take values) as
$$ Y_0 \ = \  \Psi^a_J \, \frac{\pa}{\pa u^a_J} \ . $$

Suppose that we are able to determine such a vector field which is a symmetry of $\wt{\De}$ and \emph{moreover} such that
\beql{eq:phimumu} \phi^a (x,u,w) \ = \ G^a_{\ b} (w) \ \varphi^b (x,u) \ . \eeq
Then the coefficients in the first prolongation read
$$ \Psi^a_i \ = \ D_i \phi^a \ = \ (D_i G^a_b) \, \varphi^b \ + \ G^a_b \, (D_i \varphi^b) \ . $$
As the matrix $G$ only depends on $w$, while the vector $\varphi$ only depends on $(x,u)$ we can use the decomposition \eqref{eq:D0D1a}, \eqref{eq:D0D1b}, and rewrite this -- in vector notation for ease of writing -- as
\beql{eq:psimu} \Psi_i \ = \ G \, (D_i^{(0)} \varphi) \ + \ G [G^{-1} \, (D_i^{(1)} G)] \, \varphi \ = \ G \, \[ (D_i^{(0)} \varphi ) + (G^{-1} \, D_i G) \, \varphi \] \ . \eeq
Defining the matrices $M_i$ as
$ M_i := G^{-1} \( D_i^{(1)} G \)$, i.e. as
$$ (M_i)^a_{\ b} \ = \ [G^{-1} (w)]^a_{\ c} \ \[ w^\beta_i \, \frac{\pa G^a_{\ b} (w)}{\pa w^\beta} \] \ , $$
this is also rewritten as
\beql{eq:psimu2} \Psi_i \ = \ G \, \[ (D_i^{(0)} \varphi ) + M_i \, \varphi \] \ . \eeq

Let us now take the restriction of this to the set of solutions to the auxiliary equations $\De^\beta_i$. Here $w^\beta_i = h^\beta_i (x,u,w)$, and the $w^\beta$ themselves are written in terms of the $(x,u)$ variables -- in general through expressions containing integrals of the $u^a$. We will also denote the restrictions of $G$ and $M$ to $\De^\beta_i = 0$ as
\beq \^G \ := \ \[ G \]_{\De^\beta_i = 0} \ , \ \ \ \La_i \ := \ \[ M_i \]_{\De^\beta_i = 0} \ . \eeq
Note that the $\La_i$ satisfy \eqref{eq:MCH} by construction.

With this notation, let us consider the restriction of $Y$ to the solutions of $\De^\beta_i$ and let us project it on the $J^n M$ bundle; call the resulting vector field $\^Y$. We then have
$$ \^Y \ = \ \^\psi^a_J \, \frac{\pa }{\pa u^a_J} \ , $$
where the coefficients $\^\psi^a_J$ satisfy $\^\psi^a_0 = \^\phi^a = \^G^a_b \varphi^b$ and obey the prolongation formula
\beq \^\psi^a_{J,i} \ = \ \^G^a_b \ \[ D_i^{(0)} \, \^\psi^b_J \ + \
(\La_i)^a_b \, \^\psi^b_J \] \ . \eeq
Thus, if we consider the vector field
$$ \^Z \ = \ \^G^{-1} \, \^Y \ = \ (\^G^{-1})^a_b \, \^\psi^b_J \, \frac{\pa}{\pa u^a_J} \ , $$
then this is the $\mu$-prolongation of
\beql{eq:X0mumu} X_0 \ = \ \varphi^a (x,u) \, (\pa / \pa u^a ) \eeq
for the horizontal one-form
\beql{eq:mumumu} \mu \ = \ \La_i (x,u,u_x) \, \d x^i \ . \eeq

In this case we could summarize our discussion in the form of a diagram similar to the one given above for $\la$-symmetries, i.e.
$$ \begin{matrix} \wt{\De} & \mapright{{\tt sym}} & \wt{X} & \mapright{{\tt Pr}_0} & \wt{Y} \\
\mapdown{{\tt cov}} & & \mapdown{\De^\beta_i = 0} & & \mapdown{\De^\beta_i = 0} \\
\De_0 & \mapright{\mu-{\tt sym}} & X^{(0)} & \mapright{{\tt Pr}_\mu} & Y^{(0)} \end{matrix} $$
where $\De^\beta_i = 0$ refers to the restriction to the solution manifold for the whole set of auxiliary equations $\De^\beta_i$, and we have to require that the coefficient of the $(x,u)$ variables in the vector field $X$ are as above; note that we have not discussed the functional form of the $\eta^\beta$ coefficients.\footnote{Our formulas can be slightly simplified if $G (w) = \exp [ g (w)]$; we leave this simplification to the reader.}

It is maybe convenient to summarize our discussion as a formal statement (the previous discussion gives a proof of it).

\medskip\noindent
{\bf Proposition 4.} {\it Let the system made of eqs. \eqref{eq:D0mumu} and \eqref{eq:D1mumu} -- with functions $h^\beta_i$ satisfying eq.\eqref{eq:compmumu} -- admit a Lie-point symmetry of the form \eqref{eq:Xmumu}, \eqref{eq:phimumu}. Then the equation \eqref{eq:D0mumu} admits the $\mu$-symmetry $X_0$ eq.\eqref{eq:X0mumu} with $\mu$ provided by eq.\eqref{eq:mumumu}.  }

\section{Conclusions}

We have discussed \emph{twisted symmetries}; these were introduced
as a practical tool to obtain (generalized) symmetry-reduction and
symmetry-invariant solutions for differential equations, but here
we focused on their geometrical interpretation and meaning.

In particular we considered three different approaches to them,
looking at them in different ways:
\begin{itemize}
    \item[$(a)$] consider these as standard prolongation under a
    local gauge transformation, which yields the deformed prolongation operator;
    \item[$(b)$] consider these as prolongations obtained applying
    the standard prolongation operator but with gauge-deformed
    (exterior and Lie) derivatives;
    \item[$(c)$] consider these as the image of standard prolongations
    in a covering space when projected to the original one.
\end{itemize}

It is quite clear that these different approaches are related to
each other, and we will now sketchily discuss such relations; we
hope to provide a more detailed discussion in a forthcoming work.

The approaches $(a)$ and $(b)$ are clearly and directly related, and are both based on considering gauge transformations. In the first case this is acting on vector fields which are prolonged in a standard way, i.e. on prolongation operation based on the requirement the Lie derivative of prolonged vector fields preserves the (Cartan) contact structure in $J^n M$, while in the second case the gauging is applied to the Lie derivative -- and to the exterior derivatives appearing in the contact forms -- themselves. Thus we are in a way considering  ``active'' and  ``passive'' gauging.

The relation with the approach by covering is less immediate. As we have seen, covering is based on considering new degrees of freedom (and corresponding auxiliary variables $w^\beta$), and new equations for this; the vector fields are prolonged in the standard way in the augmented phase space, but projecting this prolongation, or actually its restriction to the solutions of the auxiliary equations -- to the original space and its prolongations results in a vector field which is equivalent to a vector field prolonged by the $\la$- or $\mu$-prolongation formula.

Note that behind all of these approaches lies the basic remark -- due originally to Pucci and Saccomandi \cite{PuS} -- that twisted prolongations are vector field collinear to standard prolongations (of different vector fields), which allows them to preserve the contact structure. This is essentially due to the very basic fact that in this only the \emph{integral curves} of vector fields are relevant, and not the way the flow generated by the vector field itself runs along them.

In concrete application, one or the other of the different approaches reviewed here can be more convenient: in several cases, in particular if analyzing equations stemming from Physics, the gauge approach can yield more transparent results; on the other hand, the approach through the theory of covering makes a direct connection with \emph{non-local symmetries}, which would be quite artificial in the gauge formalism.

\section*{Acknowledgements}

The work of DCF was partially supported by the Coordena\c{c}\~{a}o
de Aperfei\c{c}\~{o}amento de Pessoal de N\'ivel Superior - Brasil
(CAPES) -  Finance Code 001, and by CNPq through grant
310577/2015-2 and grant 422906/2016-6. The work of GG was
partially supported by GNFM-INdAM. The final version of this paper
was prepared while DCF was visiting Universit\`a di Milano, thanks
to support by CAPES.



\begin{thebibliography}{99}


\bigskip

\bibitem{AVL} D.V. Alekseevsky, A.M. Vinogradov and V.V. Lychagin, {\it Basic ideas and concepts of differential geometry}, Springer 1991

\bibitem{ArnGM} V.I. Arnold, {\it Geometrical methods in the theory of differential equations}, Springer 1983

\bibitem{CF} D. Catalano Ferraioli, ``Nonlocal aspects of $\la$-symmetries and ODEs reduction'', {\it J. Phys. A: Math. Theor.} {\bf 40} (2007), 5479-5489

\bibitem{CFM1} D. Catalano Ferraioli and P. Morando, ``Local and nonlocal solvable structures in the reduction of ODEs'', {\it J. Phys. A: Math. Theor.} {\bf 42} (2009), 035210 (15pp)

\bibitem{CFM2} D. Catalano Ferraioli and P. Morando, ``Exploiting solvable structures in the integration of variational ordinary differential equations'', preprint 2014

\bibitem{Cds1} G. Cicogna, ``Reduction of systems of first-order differential equations via $\Lambda$-symmetries'', {\it Phys. Lett. A} {\bf 372} (2008), 3672-3677

\bibitem{Cds2} G. Cicogna, ``Symmetries of Hamiltonian equations and $\Lambda$-constants of motion'', {\it  J. Nonlin. Math. Phys.} {\bf 16} (2009), 43-60

\bibitem{CGbook} G. Cicogna and G. Gaeta, {\it Symmetry and perturbation theory in nonlinear dynamics}, Springer 1999

\bibitem{CGNoether} G. Cicogna and G. Gaeta: ``Noether theorem for $\mu$-symmetries'', {\it J. Phys. A} {\bf 40} (2007), 11899-11921

\bibitem{CGMor} G. Cicogna, G. Gaeta and P. Morando, ``On the relation between standard and $\mu$-symmetries for PDEs'', {\it J. Phys. A} {\bf 37} (2004), 9467-9486

\bibitem{CGWs1} G. Cicogna, G. Gaeta and S. Walcher, ``A generalization of $\lambda$-symmetry reduction for systems of ODEs: $\sigma$-symmetries'', {\it J. Phys. A} {\bf 45} (2012), 355205 (29pp)

\bibitem{CGWs2} G. Cicogna, G. Gaeta and S. Walcher, ``Orbital reducibility and a generalization of $\lambda$-symmetries'', {\it J. Lie Theory} {\bf 23} (2013), 357-381

\bibitem{CGWs3} G. Cicogna, G. Gaeta and S. Walcher, ``Dynamical systems and $\sigma$-symmetries'', {\it J. Phys. A} {\bf 46} (2013), 235204 (23pp)

\bibitem{CGWs4} G. Cicogna, G. Gaeta and S. Walcher, ``Side conditions for ordinary differential equations'', {\it J. Lie Theory} {\bf 25} (2015) 125-146

\bibitem{EGH} T. Eguchi, P.B. Gilkey and A.J. Hanson, ``Gravitation, gauge theories and differential geometry'', {\it Phys. Rep.} {\bf 66} (1980), 213-393

\bibitem{Gtwist1} G. Gaeta: ``Twisted symmetries of differential equations'', {\it J. Nonlin. Math. Phys.} {\bf 16} (2009), S107-S136

\bibitem{Gtwist2} G. Gaeta, ``Simple and collective twisted symmetries'', {\it J. Nonlin. Math. Phys.} {\bf 21} (2014), 593-627

\bibitem{Ggauge1} G. Gaeta: ``Smooth changes of frame and prolongations of vector fields'', {\it Int. J. Geom. Meths. Mod. Phys.} {\bf 4} (2007), 807-827

\bibitem{Ggauge2} G. Gaeta: ``A gauge-theoretic description of $\mu$-prolongations, and $\mu$-symmetries of differential equations'', {\it J. Geom. Phys.} {\bf 59} (2009), 519-539

\bibitem{Ggauge3} G. Gaeta, ``Gauge fixing and twisted prolongations'', {\it J. Phys. A} {\bf 44} (2011), 325203 (9 pp)

\bibitem{GFrob} G. Gaeta, ``Symmetry and Lie-Frobenius reduction of differential equations'', {\it J.Phys. A} {\bf 48} (2015)  015202

\bibitem{GMor} G. Gaeta and P. Morando, ``On the geometry of lambda-symmetries and PDE reduction'', {\it J. Phys. A} {\bf 37} (2004), 6955-6975

\bibitem{GT} J. Gibbons and S.P. Tsarev, ``Reductions of the Benney Equations'', {\it Phys. Letters A} {\bf 211} (1996), 19-24

\bibitem{God} C. Godbillon, {\it G\'eom\'etrie Diff\'erentielle et M\'ecanique Analitique}, Hermann 1969

\bibitem{GL} A. Gonzalez-Lopez, ``Symmetry and integrability by quadratures opf ordinary differential equations'', {\it Phys. Lett. A} {\bf 45} (1988), 190-194

\bibitem{KraGT} I.S. Krasil'shchik, ``A natural geometric construction underlying a class of Lax pairs'', {\it Lobachevskii J. Math.} {\bf 37} (2016), 60-65; see also {\it arXiv:1401.0612}

\bibitem{Kracov} I.S. Krasil'shchik and A.M. Vinogradov, ``Nonlocal trends in the geometry of differential equations: symmetries, conservation laws, and Backlund transformations'', {\it Acta Appl. Math.} {\bf 15} (1989),  161-209

\bibitem{KrV} I.S. Krasil'schik and A.M. Vinogradov, {\it Symmetries and conservation laws for differential equations of mathematical physics}, A.M.S. 1999

\bibitem{LNR12} D. Levi, M.C. Nucci and M.A. Rodriguez, ``$\lambda$-symmetries for the reduction of continuous and discrete equations'', {\it Acta Appl. Math} {\bf 122} (2012), 311-321

\bibitem{LR10} D. Levi and M.A. Rodriguez, ``$\lambda$-symmetries for discrete equations'', {\it J. Phys. A} {\bf 43} (2010), 292001

\bibitem{Mordef} P. Morando, ``Deformation of Lie derivative and $\mu$-symmetries'', {\it J. Phys. A} {\bf 40} (2007), 11547-11560

\bibitem{Mor14} P. Morando, ``Reduction by $\la$-symmetries and $\s$-symmetries: a Frobenius approach'', {\it J. Nonlin. Math. Phys.} {\bf 22} (2015), 47-59

\bibitem{MuRom1} C. Muriel and J.L. Romero, ``New methods of reduction for ordinary differential equations'', {\it IMA J. Appl. Math.} {\bf 66} (2001), 111-125

\bibitem{MuRom2} C. Muriel and J.L. Romero, ``$C^\infty$ symmetries and nonsolvable symmetry algebras'', {\it IMA J. Appl. Math.} {\bf 66} (2001), 477-498

\bibitem{MuRom3} C. Muriel and J.L. Romero, ``Prolongations of vector fields and the invariants-by-derivation property'', {\it Theor. Math. Phys.} {\bf 113} (2002), 1565-1575

\bibitem{MuRom4} C. Muriel and J.L. Romero, ``$C^\infty$-symmetries and integrability of ordinary differential equations''; pp. 143-150 in {\it Proceedings of the I Colloquium on Lie theory and applications (Vigo)}, 2002

\bibitem{MuRom4b} C. Muriel and J.L. Romero, ``$C^\infty$ symmetries and reduction of equations without Lie point symmetries'', {\it J. Lie Theory} {\bf 13} (2003) 167-188

\bibitem{MuRom5}  C. Muriel and J.L. Romero, ``$C^\infty$-symmetries and nonlocal symmetries of exponential type, {\it IMA J. Appl. Math.} {\bf 72} (2007) 191-205

\bibitem{MuRom6}  C. Muriel and J.L. Romero, ``Integrating factors and lambda-symmetries'', {\it J. Nonlin. Math. Phys.} {\bf 15- S3} (2008), 300-309

\bibitem{MuRom7} C. Muriel and J.L. Romero, ``First integrals, integrating factors and $\lambda$-symmetries of second-order differential equations'', {\it J. Phys. A} {\bf 42} (2009), 365207

\bibitem{MuRom11} C. Muriel and J.L. Romero, ``A $\la$-symmetry-based method for the linearization and determination of first integrals of a family of second-order ordinary differential equations'', {\it  J. Phys. A} {\bf 44} (2011), 245201

\bibitem{MuRom11b} C. Muriel and J.L. Romero, ``Second-order differential equations with first integrals of the form $C (t)+ 1/(A (t, x) x + B (t, x))$, {\it J. Nonlin. Math. Phys.} {\bf 18-S1} (2011), 237-250

\bibitem{MuRom12} C. Muriel and J.L. Romero, ``Nonlocal symmetries, telescopic vector fields and $\lambda$-symmetries  of ordinary differential equations'', {\it Symmetry, Integrability and Geometry: Methods and Applications} {\bf 8} (2012), 106-121

\bibitem{MuRom14} C. Muriel and J.L. Romero, ``The $\la$-symmetry reduction method and Jacobi last multipliers'', {\it Comm. Nonlin. Science Num. Sim.} {\bf 19} (2014), 807-820

\bibitem{Nak} M. Nakahara, {\it Geometry, Topology, and Physics}, IOP 2017

\bibitem{NS} Ch. Nash and S. Sen, {\it Geometry and Topology for Physicists}, Dover 2011

\bibitem{Olv1} P.J. Olver, {\it Application of Lie groups to differential equations}, Springer 1986

\bibitem{Olv2} P.J. Olver, {\it Equivalence, Invariants and Symmetry}, Cambridge University Press 1995

\bibitem{Pryk} A.K. Prykarpatski, ``On the Linearization Covering Technique and its Application to Integrable Nonlinear Differential Systems'', {\it SIGMA} {\bf 14} (2018), 023

\bibitem{PuS} E. Pucci and G. Saccomandi, ``On the reduction methods for ordinary differential equations'', {\it J. Phys. A} {\bf 35} (2002), 6145-6155

\bibitem{Ste} H. Stephani, {\it Differential equations. Their solution using symmetries}, Cambridge University Press 1989


\end{thebibliography}
\end{document}